\documentclass[traditabstract]{aa} % for the abstract without structuration 
\usepackage{graphicx}
\usepackage{natbib}
\usepackage{txfonts}
\newcommand{\e}{{\rm e}}
\newcommand{\ii}{{\rm i}}
\newcommand{\bA}{{{A_{\rm tot}}}}

\newcommand{\dA}{{\delta A}}
\newcommand{\dB}{{\delta B}}

\newcommand{\dS}{{\delta S}}
\newcommand{\dD}{{\delta D}}
\newcommand{\Id}{{{1}}}
\newcommand{\lap}[3]{{\rm b}_{#1}^{({#2})}({#3})}

\def \eps {\delta}

\def\abs#1{\left\vert#1\right\vert}

\def\crm{\cr\noalign{\medskip}}

\def\m@th{\mathsurround=0pt}
\def\EQM#1{\vcenter{\normalbaselines\m@th
    \ialign{${\displaystyle ##}$\hfil&&\ ${\displaystyle ##}$\hfil\crcr
    \mathstrut\crcr\noalign{\kern-\baselineskip}
    \noalign{\smallskip}
    #1\crcr\mathstrut\crcr\noalign{\kern-\baselineskip}}}}

\newcommand{\be}{\begin{equation}}
\newcommand{\ee}{\end{equation}}

\newcommand{\bpm}{\left(\begin{array}{c}}
\newcommand{\epm}{\end{array}\right)}
\newcommand{\bpmb}{\left(\begin{array}{cc}}
\newcommand{\epmb}{\end{array}\right)}
\newcommand{\figa}{
\begin{figure*}
\begin{center}
\includegraphics[width=14cm]{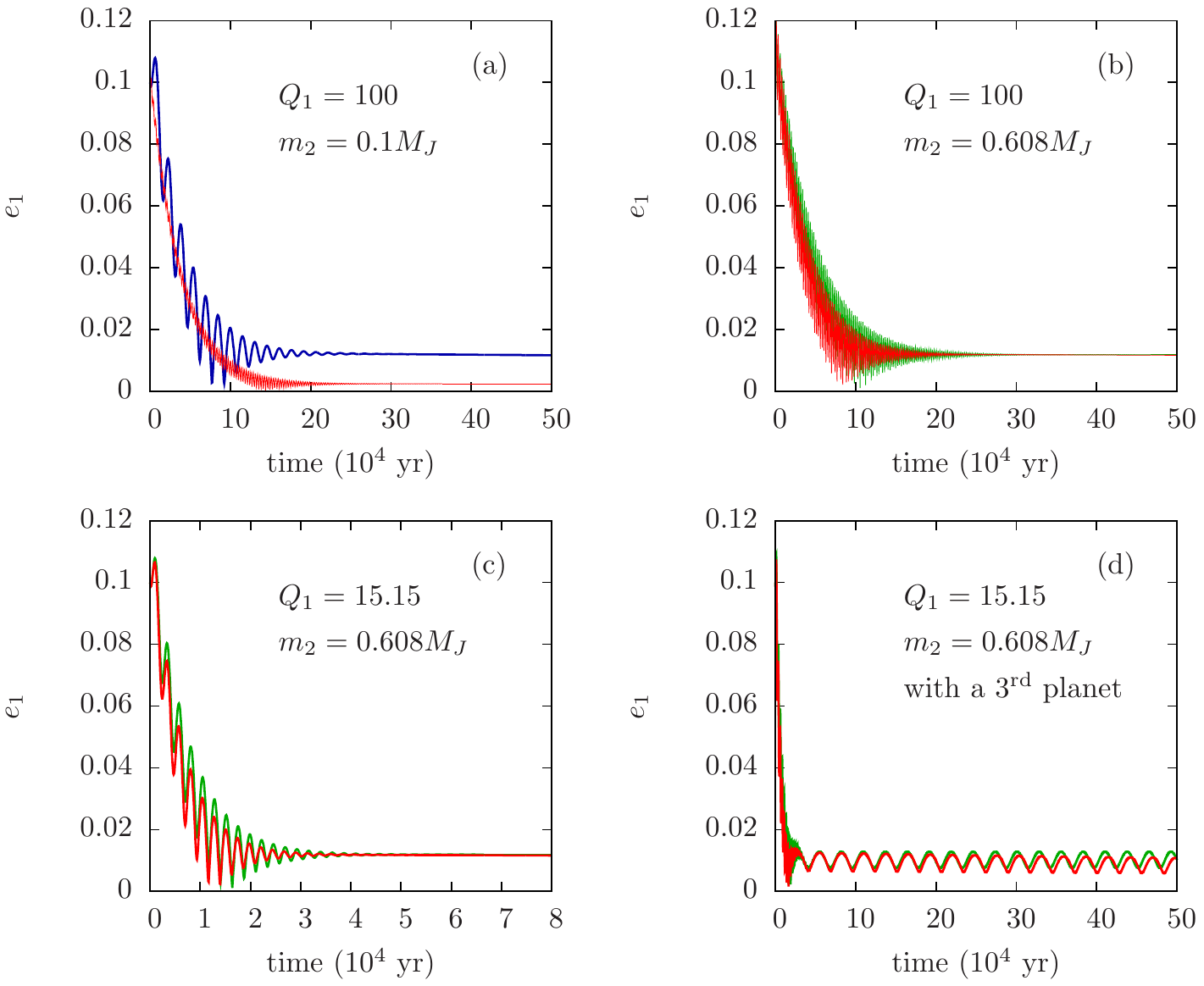}
\caption{Tidal effects on the eccentricity of HD~209458b with one or 2
companions. (a) The companion is a $0.1M_J$ planet at 0.4 AU with
$e_2=0.4$ as in \citep{Mardling_MNRAS_2007}. The blue curve has been
obtained without considering the conservative effect of the tides
($\dA^{(4)}_{11}$ in eq.~\ref{eq.tidek}). The red curve in (a), and all the
evolutions in the subfigures (b), (c), and (d) take into account this
effect. (b) The mass of the companion is set to $0.608M_J$ in order to
recover the final excentricity of Mardling's simulation. The red curve
is the result of a numerical integration of the full secular equations
exact in eccentricity. The green one is the analytical solution of the
linearized problem. (c) Same as (b) except for the initial $Q$-value of
HD~209458b which is set to 15.15 to enable the visualisation of both the
damping and the oscillation of the eccentricity. (d) Same as (c) with an
additional $0.1M_J$ companion at 1.0 AU with $e_3=0.1$.
\label{Figa}}
\end{center}
\end{figure*}
}

\newcommand{\figb}{
\begin{figure*}
\begin{center}
\includegraphics[width=14cm]{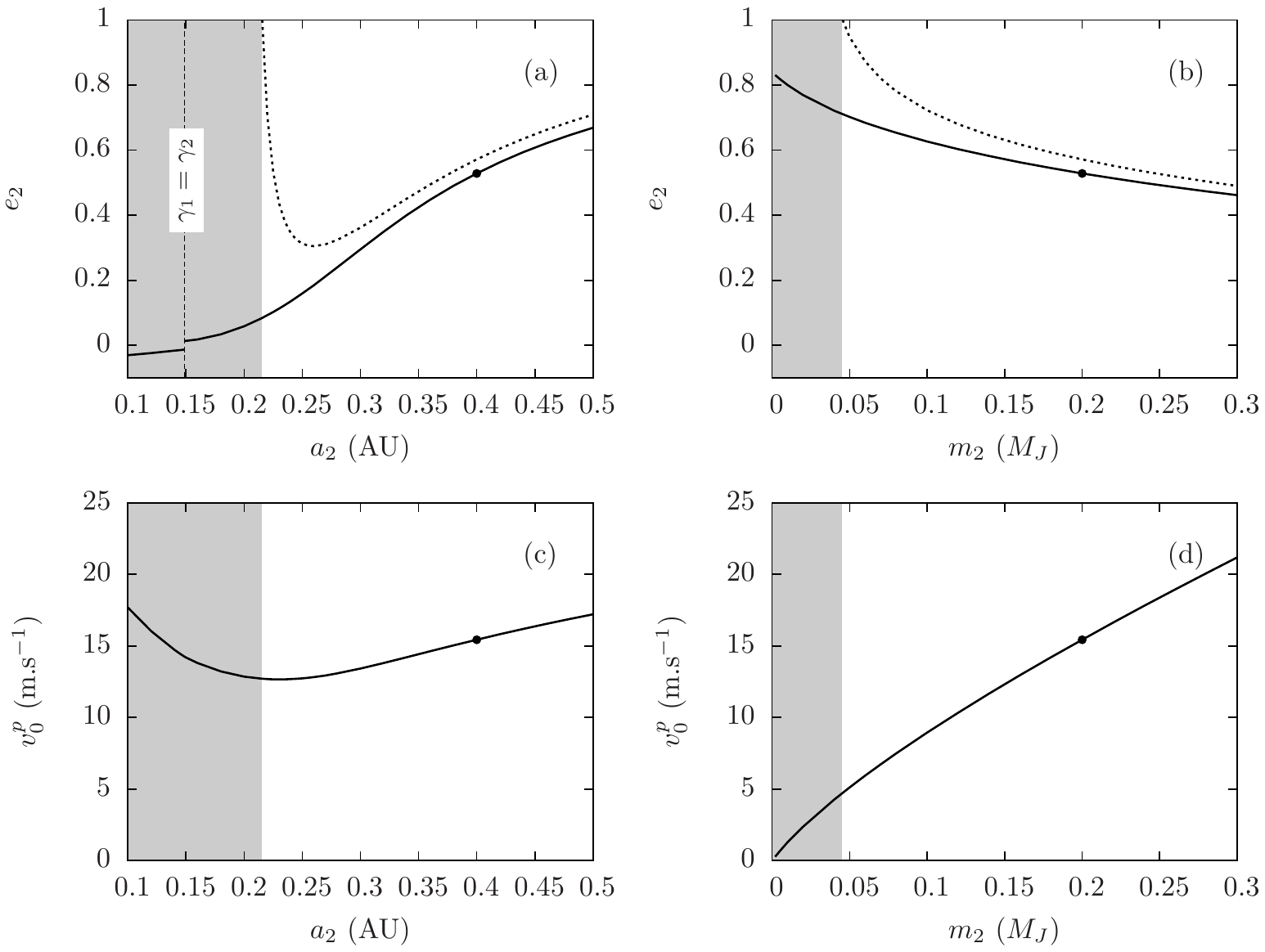}
\caption{Eccentricity $e_2$ of the hypothetical companion of HD~209458b 
with $e_1=0.01$ assuming that the eigenmode with the shortest dissipation 
timescale is damped (black curves in panel a) and b)). a) The mass of
the companion is fixed to $m_2=0.2M_J$. Negative values of $e_2$
correspond to $\Delta \varpi=180\deg$ while positive ones mean
$\Delta\varpi=0\deg$. The dotted line is the eccentricity that the
companion would have had 5.5 Gyr ago assuming a dissipation factor computed
with (\ref{eq.gammas}). b) Same as a) for different masses $m_2$ while
the semi-major axis is fixed and set to $a_2=0.4$ AU. c) Stellar reflex
velocity due to the companion at periastron with the eccentricity of the
figure a). d) idem for the eccentricity of the figure b). In grey
regions, the eccentricity of the companion should have been larger than 1
in the past. The configuration appearing in all panels with the same orbital
parameters is marked by a fill circle.
\label{Figb}}
\end{center}
\end{figure*}
}

\newcommand{\figc}{
\begin{figure}
\begin{center}
\includegraphics[width=8cm]{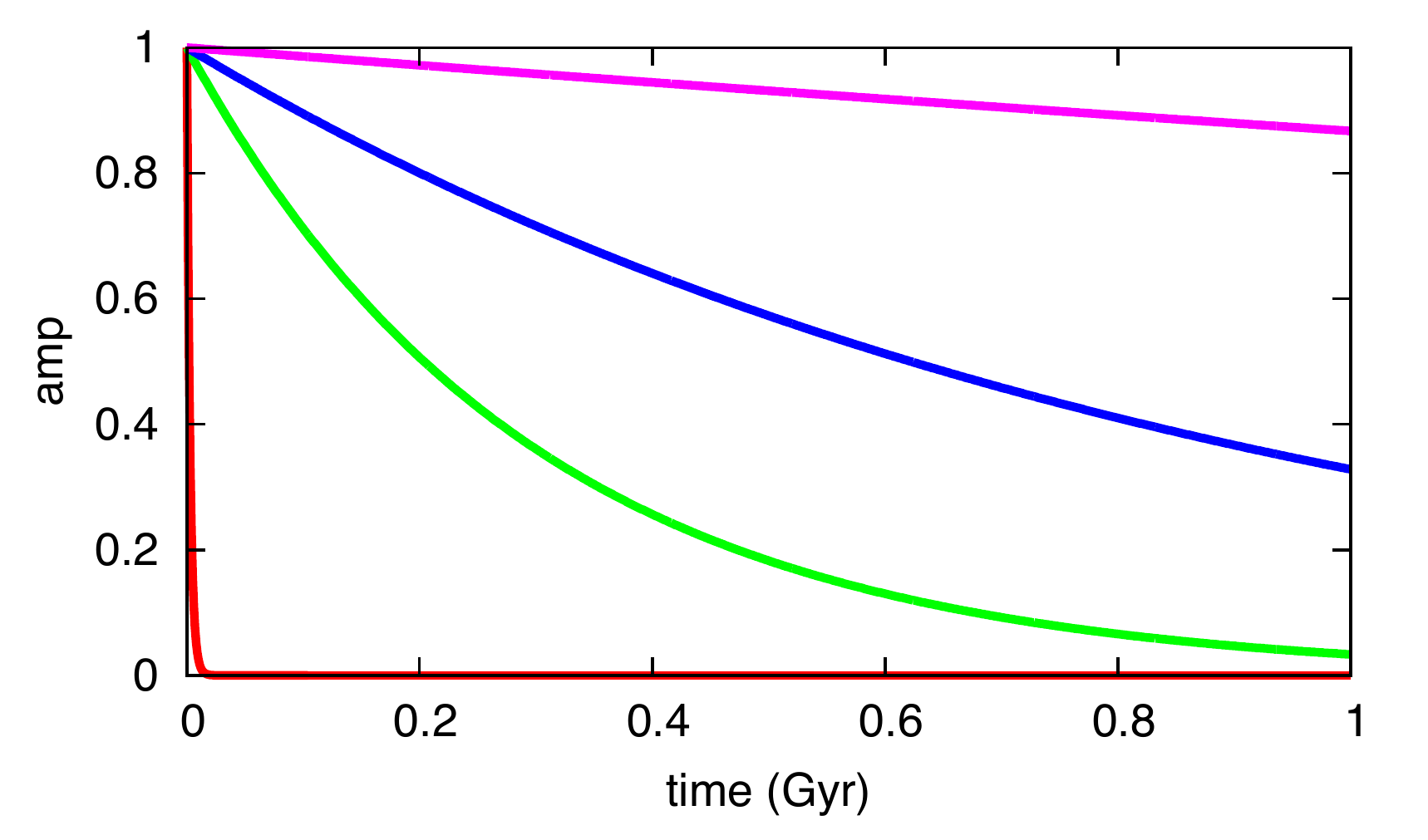}
\caption{Tidal evolution of the amplitude of the proper modes $\abs{u_1}$ (red),
$\abs{u_2}$ (green), $\abs{u_3}$ (blue), and $u_4$ (pink) resulting from 
the tidal dissipation on planet HD10180$b$ with $k_2/Q= 0.0015$ \citep{lovis_harps_2011}. 
\label{Figc}}
\end{center}
\end{figure}
}

\newcommand{\figd}{
\begin{figure}
\begin{center}
\includegraphics[width=8cm]{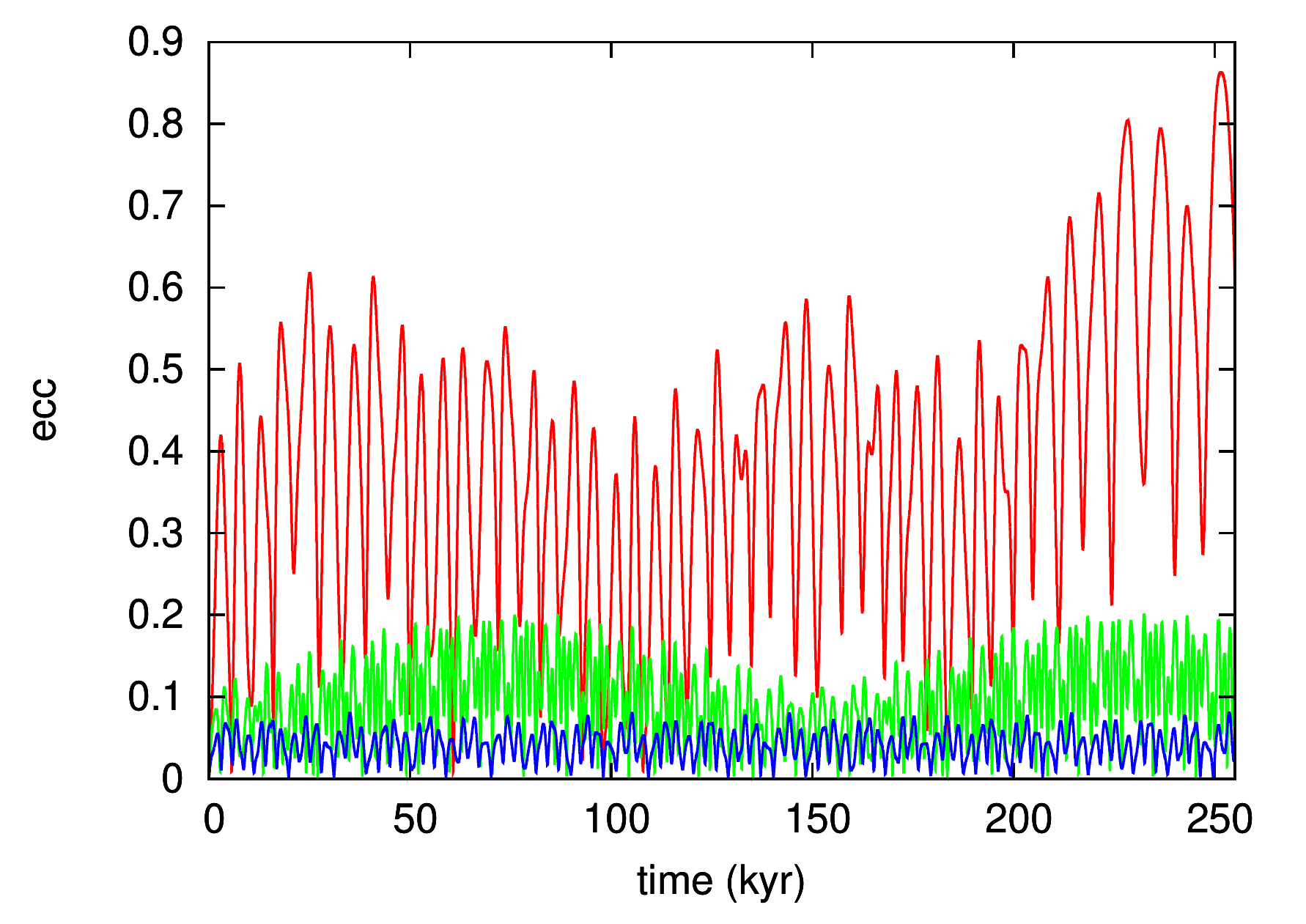}
\caption{Evolution of the eccentricity of planet HD10180b over 250 kyr starting with 
$e_b=0$ at $t=0$ (present time) for three different models : In red,  the numerical  integration 
is purely Newtonian and do not take into account general relativity (GR). In green, GR is taken into account
in the integration. In blue, GR is taken into account and the fit is made with the 
tidal dissipation constraint (\ref{eq.const}). 
\label{Figd}}
\end{center}
\end{figure}
}

\newcommand{\taba}{
\begin{table}
\begin{center}
\caption{Data for HD~209458b.}
\begin{tabular}{lc}\hline
\multicolumn{2}{c}{HD~209458b \rule[-3pt]{0pt}{13pt}} \\ \hline
Period (day) & 3.5247 \rule{0pt}{10pt} \\
$m_0$ ($M_\odot$) & 1.10 \\
$m_1$ ($M_J$) & $0.64\pm0.06$ \\
$a_1$ (AU) & 0.045 \\
$e_1$ & $0.014\pm0.009$ \\
$R_1$ ($R_J$) & $1.32\pm0.03$ \\ \hline
\end{tabular}
\label{Taba}
\end{center}
note: As in \citep{Mardling_MNRAS_2007}, all
parameters come from \citep{Burrows_etal_ApJ_2007} except the
eccentricity that comes from \citep{Laughlin_etal_ApJ_2005}.
\end{table}
}

\begin{document}
   \title{Tidal dissipation in multi-planet systems and constraints to orbit-fitting}

  % \subtitle{Tidal dissipation in multi-planet systems}

   \author{Jacques Laskar
          \inst{1}
          \and
          Gwena\"el Bou\'e\inst{1,2}
          \and
          Alexandre C. M. Correia\inst{1,3}
          }

   \institute{Astronomie et Syst\`emes Dynamiques, IMCCE-CNRS UMR8028,
Observatoire de Paris, UPMC, 77 Av. Denfert-Rochereau, 75014 Paris,
France \\
              \email{laskar@imcce.fr}
         \and
             Centro de Astrof\'\i sica, Universidade do Porto, Rua das
Estrelas, 4150-762 Porto, Portugal
             %\email{c.ptolemy@hipparch.uheaven.space}
         \and
Department of Physics, I3N, University of Aveiro, Campus Universit\'ario de Santiago, 3810-193 Aveiro, Portugal
             }

   \date{Received \ldots, 2010; accepted \ldots, \ldots}

% \abstract{}{}{}{}{} 
% 5 {} token are mandatory
 
  \abstract
   {We present here in full details the linear secular theory with tidal damping that was used 
   to constraint the fit of the HD10180 planetary system in \citep{lovis_harps_2011}. The theory 
   is very general and can provide  some intuitive understanding of the final 
   state of a planetary system when one or more planets are  close to their central star. 
   We globally recover the results of \citep{Mardling_MNRAS_2007}, 
   but we show that in the HD209458 planetary system,  
   the consideration of the tides raised by the central star on the planet 
   lead  to believe that the eccentricity of HD209458b is most probably much 
   smaller than $0.01$.}
   {}
   {}
   {}
   {}

%   \keywords{ blabla1 -- blabla2 -- blabla3 }
\titlerunning{Tidal dissipation in multi-planet systems}

   \maketitle
%
%________________________________________________________________

\section{Introduction}
In several planetary systems, some planets are very close to their central star 
and thus subject to strong tidal interaction.
If a planet were alone around its star, this will led to a circularization of its orbit. However,
for  multi-planet system, due to the  secular interaction between the planets, the final 
evolution of the system may be different, 
with residual eccentricity 
\citep{Wu_Goldreich_2002,Mardling_MNRAS_2007,Batygin_etal_2009,Mardling_MNRAS_2010,lovis_harps_2011}.  
This is also why fitting  a circular orbit to the innermost planets in a system subject to tidal 
dissipation will not insure that its eccentricity remains small,  as  the secular interactions 
may drive it to large values \citep{lovis_harps_2011}. One thus  needs  to take into account the fact 
that the observed system is the result of a tidal process  \citep{lovis_harps_2011}. Here, we develop 
in full details the method that has  been used in \citep{lovis_harps_2011} 
for the fit to a tidally evolved system. The theory is very general 
and is compared  to previous results of \citep{Mardling_MNRAS_2007,Mardling_MNRAS_2010}.

\section{Model}
\subsection{Newtonian interaction}
Without mean-motion resonances, the long term evolution of a
conservative multiplanetary system is given, in first order, by the
Laplace-Lagrange linear secular equations \citep[see][]{Laskar_Icarus_1990}.
With this approximation, inclinations and eccentricities are decoupled and
follow the same kind of evolution. For simplicity, in this letter we
will focus only on coplanar systems. Let $n$ be the number of planets.
Using the classical complex variables $z_k=e_k\e^{\ii\varpi_k}$,
$k=1,\ldots,n$, where $e_k$ and $\varpi_k$ are respectively the
eccentricity and the longitude of the periastron of the $k$-th planet,
the secular equations read
\be
\frac{d}{dt} [z]= 
\ii A\, [z]\ , \qquad \hbox{with} \quad [z]=\bpm z_1 \\ \vdots \\  z_n \epm\ ,
\label{eq.sec1}
\ee
and where $A$ is a real matrix whose elements are
\citep{Laskar_Robutel_CeMe_1995}
\be
\EQM{
A_{jj} &=& \sum_{k=1}^{j-1}n_j\frac{m_k}{m_0}C_3\left(\frac{a_k}{a_j}\right)
        +  \sum_{k=j+1}^{n}n_j\frac{m_k}{m_0}\frac{a_j}{a_k}C_3\left(\frac{a_j}{a_k}\right)
\crm
A_{jk} &=& 
\left\{
\EQM{
2n_j\frac{m_k}{m_0}\frac{a_j}{a_k}C_2\left(\frac{a_j}{a_k}\right)
\qquad & {\rm if}\quad j<k\ , \crm
2n_j\frac{m_k}{m_0}C_2\left(\frac{a_k}{a_j}\right)
\qquad & {\rm if}\quad j>k\ .\crm
}\right.
}
\ee
In these expressions, $m_k$ and $a_k$, are the mass and 
semi-major axis of the $k$-th planet, while  the mean motion  $n_k$ is defined by 
$n_k^2a_k^3=G(m_0+m_k)$. By
assumption, planets are ordered by increasing semi-major axis
while the index 0 stands for the
star. The functions $C_2(\alpha)$ and
$C_3(\alpha)$ are defined by mean of ${\rm b}_s^{(k)} $, the usual
Laplace coefficients \citep[e.g.][]{Laskar_Robutel_CeMe_1995}, as:
\be
\EQM{
C_2(\alpha) &=& \frac{3}{8}\alpha\lap{3/2}{0}{\alpha}
-\left(\frac{1}{4}+\frac{1}{4}\alpha^2\right)\lap{3/2}{1}{\alpha}\ ,\crm
C_3(\alpha) &=& \frac{1}{4}\alpha\lap{3/2}{1}{\alpha}\ .
}
\ee

\subsection{Other effects}
The above secular equations (\ref{eq.sec1}) describe only the Newtonian
interactions between point mass planets. In order to study exoplanetary
systems with short period planets, it is often necessary to add
corrections due to relativity, oblateness and tidal friction, at least
to the inner planets. The spatial secular equations of motion resulting
from all these effects are given in \citep{Lambeck_1980,Eggleton_Kiseleva_ApJ_2001,Ferraz-Mello_etal_2008}. To
first order in eccentricity, and in the planar case, they modify the
diagonal terms of the matrix $A$ in two different ways. There are
conservative terms that are purely imaginary, and dissipative ones which
are real (and negative). We thus define two new diagonal matrices
$\delta A = \sum_{i=1,5}\delta A^{(i)}$ and
$\delta B=\sum_{i=4,5}\delta B^{(i)}$ such that the full secular evolution is given by
\be
\frac{d}{dt}[z] = 
(\ii\bA-\dB)[z]\ ,
\label{eq.sec2}
\ee
with $\bA=A+\dA$. The effect of relativity on the
$k$-th planet is conservative, and in first order in eccentricity, it
leads to
\be
\delta A^{(1)}_{kk} = 3\frac{Gm_0}{c^2}\frac{n_k}{a_k}\ .
\label{eq.rel}
\ee
The effect of the oblateness of bodies generated by their proper rotation are
\be
\delta A^{(2)}_{kk} = \frac{k_{2,k}\,n_k}{2}\left(\frac{m_0+m_k}{m_k}\right)
            \left(\frac{R_k}{a_k}\right)^5\left(\frac{\omega_k}{n_k}\right)^2\ ,
\ee
and
\be
\delta A^{(3)}_{kk} = \frac{k_{2,k}\,n_k}{2}\left(\frac{m_0+m_k}{m_0}\right)
            \left(\frac{R_0}{a_k}\right)^5\left(\frac{\omega_0}{n_k}\right)^2\ ,
\ee
for the oblateness of the $k$-th planet and the oblateness
of the star.
$k_{2,k}$, $\omega_k$, and $R_k$ are respectively
the second  Love number, the proper
rotation rate, and the radius of the $k$-th body.
Tidal effects have two contributions. With the same
approximation, we have
\be
\delta A^{(4)}_{kk} = \frac{15}{2}K_k\ ,\qquad
\delta B^{(4)}_{kk} = 27\left(1-\frac{11}{18}\frac{\omega_k}{n_k}\right)\frac{K_k}{Q_k}\ ,
\label{eq.tidek}
\ee
with
\be
K_k = k_{2,k}n_k
\left(\frac{m_0}{m_k}\right)\left(\frac{R_k}{a_k}\right)^5
\ee
for the tides raised on the $k$-th planet by the star, and
\be
\delta A^{(5)}_{kk} = \frac{15}{2}K'_k\ ,\qquad
\delta B^{(5)}_{kk} = 27\left(1-\frac{11}{18}\frac{\omega_0}{n_k}\right)\frac{K'_k}{Q_0}\ ,
\label{eq.tide0}
\ee
with
\be
K'_k = k_{2,0}n_k
\left(\frac{m_k}{m_0}\right)\left(\frac{R_0}{a_k}\right)^5
\ee
for the tides raised on the star by the $k$-th planet. 
We consider here the ``viscous'' approach  \citep{Singer_1968,Mignard_MP_1979}, where
the quality factor of the $k$-th body is
$Q_k\equiv (n_k(\Delta t)_k)^{-1}$ and $(\Delta t)_k$ is a constant time
lag. 
\section{Resolution}
\subsection{Conservative case}
When there is no dissipation ($\delta
B=0$) the system is classically resolved by diagonalizing
the matrix $\bA$ through a linear transformation 
\be
[z]=S_0 [u] \ .
\label{eq.chvar0}
\ee
In the new variables, the equations of
motion become
\be
\frac{d}{dt}[u] = \ii D_0[u] \ ,\quad \hbox{where} \quad  D_0=S_0^{-1}\bA S_0
\label{eq.D0}
\ee
is the diagonal matrix ${\rm diag}(g_1,\ldots,g_n)$ of the 
eigenvalues $g_k$ of $\bA$. We have then
\be
u_k(t) = u_k(0)\e^{\ii\, g_k t}\ .
\ee
Each proper mode $u_k$ describes a circle in the complex plane at 
a constant frequency $g_k$ and with the radius $\abs{u_k(0)}$.
The evolution of the planetary
eccentricities are then given by (\ref{eq.chvar0}). They are linear
combinations of the proper modes.

The only differences between $A$ and $\bA$ are in the diagonal terms.
Those of $\bA$ are larger  or equal to  those of $A$. As a consequence,
using $\bA$ instead of $A$ makes the fundamental frequencies $g_k$
larger and the coupling between the proper modes lower (the evolution of
the eccentricity of each planet is almost given by one single proper mode,
the other modes generate only small oscillations).

\subsection{General solution}
\label{sec.gen}
In the full linear secular equation (\ref{eq.sec2}), the matrix that has
to be diagonalized is now $\ii\bA-\dB$, where the dissipation part $\dB$
comes only from tides. In general, the elements of $\dB$ are much smaller than
those of the diagonal of $\bA$. $\dB$ will thus  be
considered as a perturbation of the conservative evolution given by $\bA$. 
Let
\be
S=S_0 ( \Id +\ii\eps S_1 ) \ ,
\label{eq.S}
\ee
be the matrix of the linear transformation that diagonalizes the full
system. As $\dB$ is a perturbation of $\bA$, 
we will make the hypothesis that 
the matrix $\dS_1$ is also
a perturbation of the matrix $S_0$. At first order, the inverse of $S$ is
\be
S^{-1} = (\Id-\ii\dS_1)S_0^{-1}\ ,
\ee
and the new diagonal matrix is $D=iD_0-\dD_1$, with
\be
\dD_1 = S_0^{-1}(\dB)S_0 - [\dS_1,D_0]\ ,
\label{eq.D1}
\ee
where the bracket is defined by
$ [\dS_1,D_0]=\dS_1 D_0 - D_0 \dS_1$. 
For $\dD_1$ to be actually diagonal, $\dS_1$ is given
by
\be
(\dS_1)_{jk} =
\frac{1}{g_k-g_j}\big(S_0^{-1}(\dB)S_0\big)_{jk}
\ ,\qquad j\neq k\ .
\ee

As $D_0$ is diagonal, all terms in the diagonal of $[\dS_1,D_0]$ vanish.
Thus, the diagonal terms of $\dS_1$ do not appear in the
computation of $\dD_1$ (\ref{eq.D1}) and they can be set equal to zero.
Let $\dD_1={\rm diag}(\gamma_1,\ldots,\gamma_n)$. From (\ref{eq.D1}), we
have then
\be
\gamma_k = \big(S_0^{-1}(\dB)S_0\big)_{kk}\ .
\label{eq.gamma}
\ee
These $\gamma_k$ are real and positive. It turns out that the imaginary
part of $D$ is still the one of the conservative case $D_0$
(\ref{eq.D0}). The proper frequencies $g_k$ are not affected by the
dissipation $\dB$. However, each proper mode now contains a damping
factor $\gamma_k$ given by (\ref{eq.gamma}).  The equations of
motion in the new variables now read
\be
\frac{d}{dt}[u] = 
{\rm diag}(\ii g_1-\gamma_1,\ldots,\ii g_n-\gamma_n)
[u]\ ,
\ee
and the solutions are
\be
u_k(t) = u_k(0)\e^{-\gamma_kt}\e^{\ii g_kt}\ .
\label{eq.u}
\ee

It should be stressed that even if only one planet undergoes tidal
dissipation (only $(\delta B)_{1 1}$ is different from 0 for example),
because of the linear transformation $S_0$, all the eigenmodes can be
damped (\ref{eq.gamma}).

\section{Two planet case}
In a simpler two planet system  where
only the first one undergoes tidal friction, $\dB ={\rm diag}(\gamma,0)$,
the two proper frequencies are given by
\be
\EQM{
g_1 &=& \frac{1}{2}\left(T+\sqrt{T^2-4\Delta}\right)\ , \crm
g_2 &=& \frac{1}{2}\left(T-\sqrt{T^2-4\Delta}\right)\ ,
}
\ee
where $T$ and $\Delta$ are the trace and determinant of $A_{tot}$.
From (\ref{eq.gamma}), it can be shown that the two dissipation factors
are
\be
\EQM{
\gamma_1 &=& \frac{1}{2}\left(1+\frac{A_{11}-A_{22}}{g_1-g_2}\right)\gamma
\ ,\crm
\gamma_2 &=& \frac{1}{2}\left(1-\frac{A_{11}-A_{22}}{g_1-g_2}\right)\gamma
\ .
}
\label{eq.gammas}
\ee
The sum $\gamma_1+\gamma_2$ is equal to $\gamma$. There is thus always
one eigenmode damped in a timescale shorter than $2\gamma^{-1}$ while
the other is damped in a timescale larger than $2\gamma^{-1}$. In the
particular case where $A_{11}=A_{22}$, we have
$\gamma_1=\gamma_2=\gamma/2$.

Once the first eigenmode is damped, the ratio between the two
eccentricities and the difference between the two longitudes of
periastron are deduced from (\ref{eq.chvar0}). We have
\be
\EQM{
\frac{e_1}{e_2} &=& \sqrt{\frac{A_{12}}{A_{21}}\frac{\gamma_2}{\gamma_1}}&
\ ,\quad \varpi_1-\varpi_2=0\ ,\quad&
{\rm if}\ \gamma_1>\gamma_2\ , \crm
\frac{e_1}{e_2} &=&\sqrt{\frac{A_{12}}{A_{21}}\frac{\gamma_1}{\gamma_2}}&
\ ,\quad \varpi_1-\varpi_2=\pi\ ,\quad&
{\rm if}\ \gamma_1<\gamma_2\ .
}
\label{eq.erat}
\ee
It should be noted that the matrix $\dS_1$ introduces small corrections
in the difference between the longitudes of periastron which are not
taken into account in (\ref{eq.erat}).

\figa

\figb

\section{Application to HD~209458b}
Here we compare the results of this paper with those of
\citet{Mardling_MNRAS_2007} on the example of HD~209458b
(Table~\ref{Taba}). As in \citep{Mardling_MNRAS_2007}, we first assume that
the non zero eccentricity of this planet is due to the presence of a
$m_2=0.1M_J$ companion at $a_2=0.4$ AU with an eccentricity $e_2=0.4$.
For this study, eccentricities are large and modify the frequencies
$g_k$ given by the analytical expression of the matrix $\bA$
(\ref{eq.sec2}). Thus, we chose to compute the matrix $\bA$ using a
frequency analysis on a numerical integration of the system without
dissipation exact in eccentricity and expanded up to the 4 order in the
ratio of the semi-major axes
\citep[e.g.][]{Mardling_Lin_ApJ_2002,Laskar_Boue_AA_2010}. At first order, the
eccentricity variables $z_1$ and $z_2$ are linear combinations of two
eigenmodes $u_1$ and $u_2$ (\ref{eq.u})
\be
\EQM{
z_1(t) &=& S_{11}u_1(t) + S_{12}u_2(t)\ , \cr
z_2(t) &=& S_{21}u_1(t) + S_{22}u_2(t)\ ,
}
\label{eq.coupled}
\ee
where $S$ is given by (\ref{eq.S}). With $Q_1=10^5$ and $\omega_1=n_1$,
the two damping timescales (\ref{eq.gammas}) are $\gamma_1^{-1}\sim\gamma^{-1}=46$ Myr and
$\gamma_2^{-1}=589$ Gyr. 
%With an age estimate of 4.5 Gyr for this system \citep{Mazeh_etal_ApJ_2000}, 
With an age estimate of 5.5 Gyr for this system \citep{Burrows_etal_ApJ_2007}, 
the first eigenmode should be damped and
the modulus of the second should remain almost constant. In consequence,
both eccentricity variables should be proportional to $u_2$. Their modulus
should thus be constant and verify (\ref{eq.erat}), or equivalently, 
$e_1 \approx  e_2\,S_{12}/S_{22}
= 0.0025$. In \citep[][Fig.~3]{Mardling_MNRAS_2007}, this value is larger,
$e_1=0.012$. The difference comes from the tidal deformation of the planet that leads
to the coefficient $\dA^{(4)}_{11}$ in Eq.~(\ref{eq.tidek}). This was not
taken into account in \citep{Mardling_MNRAS_2007}, and it accelerates
the precession of the periapse of HD~209458b by a factor 6.6 (see
Fig.~\ref{Figa}a). In figure~\ref{Figa}a, the initial $Q$-value of the
planet is set artificialy to 100 to enable a direct comparison with the
figure~3 of \citep{Mardling_MNRAS_2007}. As said by
\citet{Mardling_MNRAS_2007}, and showed in this paper, the $Q$-value
only affects the damping timescales but not the precession frequencies,
nor the eccentricity amplitudes. However, with a larger precession
frequency, the matrix $\bA$ is closer to a diagonal matrix. The two
planets are less coupled and the ratio $S_{12}/S_{22}$
(\ref{eq.coupled}), equal to the final eccentricity ratio $e_1/e_2$, is
smaller.

\taba

One way to recover the final eccentricity of HD~209458b is to increase
the mass of the companion up to $m_2=0.608M_J$ (Fig.~\ref{Figa}b).
Here, our aim is not to explain the large eccentricity of HD~209458b,
but simply to illustrate the results of the section \ref{sec.gen}.

As the precession of the periastron of the inner planet is faster than
in \citep[][Fig.~3]{Mardling_MNRAS_2007}, we decreased the initial $Q$-value to
15.15 to accelerate the damping and to obtain an evolution with the same
$g_1/\gamma_1$ ratio as in \citep{Mardling_MNRAS_2007}
(Fig.~\ref{Figa}c). As said before, this does not change the final
eccentricity, but it illustrates better the damping of the first mode
with frequency $g_1=0.14$ deg/yr.

After the damping of the first eigenmode, the eccentricities are not
oscillating because there remains only one eigenmode with a non-zero
amplitude. Both eccentricity variables $z_1$ and $z_2$ describe a circle
in the complex plane at the same frequency $g_2$. But if a third planet
is added to the system, a new eigenmode appears with a frequency $g_3$. 
Then eccentricities are oscillating (Fig.~\ref{Figa}d). It should be
noted that a relative inclination between planets can also generate an
other eigenmode and make eccentricities oscillate
\citep{Mardling_MNRAS_2010}. However, in the linear approach
eccentricities and inclinations are decoupled. It is thus necessary to
have large eccentricities or inclinations to have significant
oscillations.

We now wonder which companion parameters can lead to an eccentricity
$e_1=0.01$ for HD~209458b. As the system contains two planets,
eccentricities are at most combination of two eigenmodes. But since
$\gamma^{-1}=46$Myr is less than the age of the system (5.5 Gyr), at
least one of the eigenmode is damped.  However both eigenmodes cannot
have zero amplitude, else the two orbits would be circular. Thus, let us
assume that  remains only a single eigenmode, the one with the
longer damping timescale. Then, given a semi-major axis $a_2$ and a mass
$m_2$, the eccentricity of the companion is obtained through
(\ref{eq.erat}) at first order. In practice we integrated numerically
the system without dissipation, and found the eccentricity $e_2$ that
cancels the amplitude of the rapidly damped eigenmode. Results are shown
with solid curves in figure \ref{Figb}a and \ref{Figb}b. Once the
current eccentricity $e_2$ is given, the initial value (5.5 Gyr ago) is
estimated assuming an exponential decay with a damping factor given by
(\ref{eq.gammas}) (see the dotted curves Fig.~\ref{Figb}a and
Fig.~\ref{Figb}b). The frequencies $g_k$ and the coefficients $A_{kk}$
were obtained numerically using a frequency analysis. Parameters leading
to initial eccentricities larger than 1 are excluded. They correspond to
the grey regions in figure \ref{Figb}. Although planets are less coupled
than in \citep{Mardling_MNRAS_2007}, there is still a large range of
initial conditions leading to a state compatible with $e_1=0.01$. However,
the stellar reflex velocity due to the companion at periastron
(Fig.~\ref{Figb}c and \ref{Figb}d) is above the detectability
threshold of about 3 m$.$s$^{-1}$. For example, with $a_2=0.25$ AU, and
$m_2=0.05M_J$, the current eccentricity is $e_2=0.34$ and the maximal 
stellar reflex velocity $v_0^p=3.9$ m$.$s$^{-1}$. 
% In that case, the current initial eccentricity should be 0.74.
It thus seems that such a planet cannot exist. Indeed, observations
do not constrain strongly the eccentricity of HD~209458b and a circular
orbit is not ruled out \citep{Laughlin_etal_ApJ_2005}.

%__________________________________________________________________

\section{Orbit fitting : the HD10180 case}

\figc
\figd

The analysis of the radial velocities measures of HD10180 revealed the potential existence of 
7 planets in this system \citep{lovis_harps_2011}. The innermost planet,  
HD10180b, is a terrestrial planet ($m_b\sin i = 1.35 M_\oplus$)
with   period of $\approx 1.177$ days and  semi-major axis $a_b= 0.0223$ AU. 
The planet is thus   subject to strong tidal interaction with the central star. 
During the first fit \citep[Table 3]{lovis_harps_2011} , it was thus assumed that its eccentricity has been damped 
to very small values, and its value was fixed to $e_b=0$. 
Nevertheless, if the system is then numerically integrated  over 250 kyr (Fig. \ref{Figd} (red curve)), 
due to secular interactions with the other planets, $e_b$ grows very rapidly to high 
values, reaching nearly 0.9.

When general relativity (GR) is included in the numerical integration, the main effect is to 
increase the diagonal terms  of the secular matrix (Eq.\ref{eq.rel}). As a result, 
the secular variations of $e_b$ are much smaller (Fig. \ref{Figd} (green curve)), but still 
reach   $0.2$. 

The strategy that was adopted for the final fit of \citep{lovis_harps_2011} was to include in the fit 
the constraint that the planetary system that is searched for is the result 
of  the tidal evolution, as described in section \ref{sec.gen}. As the planet 
has a mass comparable to the Earth, it can be assumed to be  terrestrial, 
and thus to have a dissipation factor   of the same  order of 
magnitude  as (or larger than) Mars $k_2/Q = 0.0015$, which is the smallest value among the 
terrestrial planets  in the Solar System. 
The damping factors $\e^{-\gamma_k}\,t$  can thus be computed 
through (Eq. \ref{eq.gamma}) for all proper modes $u_k$ \citep[Table 5]{lovis_harps_2011}. 
The resulting dampings of the  amplitudes of the proper modes $u_k$ are given in 
Fig.\ref{Figc}.

From this computation, as the age of the system is estimated to be of about 4 Gyr, 
it can be seen that the first two proper modes amplitudes $u_1$ and $u_2$  are certainly 
reduced to very small values. If the damping factor $k_2/Q$ were 10  times smaller, 
the conclusion would be nearly the same, as the only change in Fig.\ref{Figc} would be 
a change in the time scale of the figure, the units being now 10 Gyr instead of 1 Gyr.

In order to include the constraint on the tidal damping in the fit, one can then 
add to the $\chi^2$ minimization the additional term 

\be
\chi^2_R = R(\abs{u_1}^2 + \abs{u_2}^2) 
\label{eq.const}
\ee
where $R$ is an empirical  constant that needs to be set to a value that will 
equilibrate the additional constraint with respect to the 
value of the $\chi^2$ in absence of constraint. 
After various trials, $R=350$ was used in \citep{lovis_harps_2011}.

The computation of the amplitude of the proper modes $\abs{u_k}$ during the fit 
is made iteratively. Once a first orbital solution is obtained, the Laplace-Lagrange 
linear system (Eq.\ref{eq.sec1}) is computed and thus also the matrix $S_0$ 
of transformation  to proper modes  (Eq.\ref{eq.chvar0}). For a given 
initial condition $(z_k)$ obtained through the fit, the proper modes $u_k$ are computed with 
\be
[u] = S_0^{-1}
[z]
\label{eq.chvar0m1}
\ee
and the additional contribution (\ref{eq.const})
can then be computed in the fitting process. 
Practically, in an iterative fit  taking into account the 
Newtonian interactions, the  transformation matrix $S_0^{-1}$  just needs to be computed 
once, or twice if one wants to recompute the $S_0^{-1}$ matrix 
when the convergence to a final solution is obtained. 
In \citep{lovis_harps_2011},  the final values were $u_1= 0.0017$, $u_2=0.044$ 
for $R=350$, with a final $\sqrt{\chi^2}=1.24$, very close to the residuals 
obtained in absence of constraint ( $\sqrt{\chi^2}=1.22$).

In this constrained solution, the initial value of $e_b$ is still $0$, 
but the secular change due to planetary interactions is much smaller
(Fig.\ref{Figd} (blue curve)),  which ensure a more stable behavior to the system.

%__________________________________________________________________

\section{Conclusion}

We have presented here in full details the secular theory with 
tidal dissipation that was outlined in \citep{lovis_harps_2011} for the system HD10180. 
The use of Lagrange-Laplace linear theory can include very easily the 
linear contribution due to tidal dissipation and provide 
an intuitive background for  studying  multi-planetary systems 
when one or several planets are close to their central star and 
subject to  tidal damping. Although  we have limited here the study to the planar case,  
this formalism can be easily  extended   to 
mutually inclined systems.

For the system HD209458,
we could retrieve globally the results of \citep{Mardling_MNRAS_2007}, 
although we find that  a companion with mass $m_2 = 0.1 M_J$  with $a_2=0.4$ 
AU and $e_2=0.4$  will not lead to $e_1 = 0.012$, but to a much smaller value 
of $e_1 = 0.0025$. This is  due to the additional 
tides raised by the star on the planet $\delta A_{11}^{(4)}$ (Eq.\ref{eq.tidek}) 
in the contribution to the secular equations  
(Eq.\ref{eq.sec2}).

We have examined other configuration that could lead to a final eccentricity $e_1\geq 0.01$ 
for HD209458b, but our conclusion is negative, as we found that  a potential 
companion, large enough to lead to a final eccentricity $e_1 \geq 0.01$ leads 
to sufficiently large stellar motion that it should have already been detected, 
assuming a detectability threshold of 3 m.s${}^{-1}$.
Our conclusion is thus that the most probable outcome is that the actual eccentricity 
of HD209458b has a much smaller value than $0.01$. 

\section*{Acknowledgments}
This work has been supported by PNP-CNRS, by the European Research Council/European
Community under the FP7 through a Starting Grant, as well as in the form
of grant reference PTDC/CTE-AST/098528/2008, funded by Funda\c{c}\~ao
para a Ci\^encia e a Tecnologia (FCT), Portugal.

\bibliographystyle{aa}
\bibliography{damping}
\end{document}